\documentclass[11pt]{elsarticle}
\usepackage{amsmath,amssymb}
\usepackage[margin=2.5cm]{geometry}
\usepackage{graphics,graphicx}
\usepackage{dcolumn,bm}
\usepackage{psfrag}
\usepackage{color}
\newcommand{\be}{\begin{equation}}
\newcommand{\ee}{\end{equation}}

\usepackage{graphics,graphicx}
\journal{Physica A}

\begin{document}
\title{The Ising universality class of kinetic exchange models of opinion dynamics}

\author{Sudip Mukherjee${}^{a,b,}$\footnote{sudip.mukherjee@saha.ac.in}}
\author{Soumyajyoti Biswas${}^{c,}$\footnote{soumyajyoti.b@srmap.edu.in}}
\author{Arnab Chatterjee${}^{d,}$\footnote{arnab.chatterjee4@tcs.com}}
\author{Bikas K. Chakrabarti${}^{b,e,f,}$\footnote{bikask.chakrabarti@saha.ac.in}}
\address{${}^a$Department of Physics, Barasat Government College, Barasat, Kolkata 700124, India\\
${}^b$Saha Institute of Nuclear Physics, Kolkata 700064, India \\
${}^c$Department of Physics, SRM University - AP, Andhra Pradesh - 522502, India\\
${}^d$TCS Research, Tata Consultancy Services, New Delhi, India \\
${}^e$S. N. Bose National Centre for Basic Sciences, Kolkata 700106, India \\
${}^f$Economic Research Unit, Indian Statistical Institute, Kolkata 700108, India
}

\date{\today}

\begin{abstract}
We show using scaling arguments and Monte Carlo simulations that a class of 
binary interacting models of opinion evolution belong to the Ising universality class in 
presence of an annealed noise term of finite amplitude. While the zero noise limit is known to show 
an active-absorbing transition, addition of annealed noise induces a continuous order-disorder
transition with Ising universality class in the infinite-range (mean field) limit of the models. 
\end{abstract}

\maketitle

\section{Introduction}

Study of social dynamics using tools of statistical physics is an active research area, where 
systems and models are being studied qualitatively and 
quantitatively~\cite{stauffer2006biology,Chakrabarti2006econosocio,CFL_RMP2009,
galam2012sociophysics,Sen:2013}. The study of dynamics of opinions, individual and 
collective choices is one of the most popular topics.
One is usually concerned with the question, how interacting individuals choose between
different options (vote,  opinions, language, culture, etc.), leading
to a state of `consensus' or a state of coexistence of multiple options.
With the basic assumption that such options can be quantified and that the individuals 
take their decisions in choosing those options through interactions with their `neighbors',
physicists have attempted to model this `complex', multi-component system to bring forward
their generic and universal features \cite{CFL_RMP2009,galam2012sociophysics,Sen:2013}.
  
Given a finite interaction term among the individuals, that competes with the noise, usually the models of social dynamics exhibit collective 
dynamical phenomena. Several models introduced 
so far study the dynamics that leads to different opinion states and the universal nature of 
 transitions between such states.
The rich emergent phenomena, resulting out of interaction of a large 
number of entities or agents~\cite{liggett1999stochastic} follow  various `universality classes' in terms of their characterizations that are, depending on specific contexts, 
similar to the models of statistical physics~\cite{CFL_RMP2009,Sen:2013}.

The present study concerns the dynamics of opinions, and how consensus 
may or may not emerge out of interaction of individual opinions, or, out of interaction of 
individuals with opinions evolving out of influence of others. 
A series of studies on this 
topic~\cite{lewenstein1992statistical,deffuant2000mixing,hegselmann2002opinion,
galam1982sociophysics,galam2002minority} has enriched our understanding of the topic.
Opinions are usually modeled as discrete or continuous variables, which can undergo 
spontaneous changes or change due to interaction with others, or external factors. 
The study of dynamics of opinions, as well as their steady state properties are interesting. 
The possibility of a phase with a spectrum of opinions and another phase where the majority 
have similar values may demonstrate the existence of different phases -- a typical scenario to 
study phase transitions. In continuous opinion models, opinions cluster around a single value 
(consensus), or two (polarization) or can as well have several values (fragmentation).

We focus our attention to a specific class of models ~\cite{lallouache2010opinion}, having apparent similarity with kinetic models of 
wealth exchange~\cite{Chatterjee2007,chakrabarti2013econophysics}. The opinions of individuals that can be continuous or discrete variables, depending on 
the context, are bound between two extremes $(-1,1)$. The agents update their opinions, at least in the leading order,
 due multiple binary interactions between them. In this and in a set of variants of the model, there is no external noise term 
(arising out of, say, imperfect information received by the agents) \cite{sen2011phase,biswas2011phase,biswas2011mean}.   
Invariably, in those models, active-absorbing transitions between an indifferent state and that with a dominant variant of opinion in the society were seen. 
In other variants, disorder/noise were introduced in the binary interactions, which led to 
order to disorder type with Ising exponents~\cite{biswas2012disorder,BCS2d3d}.

Here we show that the steady state statistics of two kinetic exchange models of opinion formation have Ising universality class. 

\begin{figure}[t]
\includegraphics[width=17.5cm]{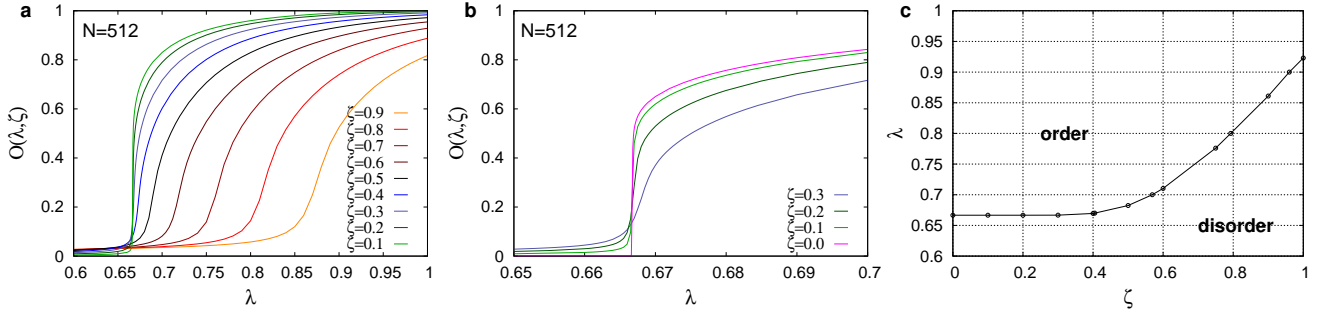}
\caption{Plot of order parameter with conviction $\lambda$ for different values of noise: 
(a) for $\zeta=0.1,0.2,0.3,0.4,0.5,0.6,0.7,0.8,0.9$, and (b) for $\zeta=0.1,0.2,0.3$ in a 
smaller range of $\lambda$, compared with the LCCC model ($\zeta=0$). The plots are given for 
$N=512$.
(c) The phase diagram of the noisy LCCC model, in the $\lambda-\zeta$ plane.
The region above the line is the `ordered' phase while that below is the `disordered' phase.
 }
 \label{fig:nlc3M}
\end{figure}

\section{Kinetic exchange opinion model: Transition driven by external noise}\label{sec:2}
Following the multi-agent statistical model
of closed economy \cite{Chatterjee2007}, Lallouache et al.~\cite{lallouache2010opinion} 
proposed a minimal multi-agent model for the collective dynamics of opinion  formation. 
Let $o_i(t) \in (-1,+1)$ be the opinion of an individual $i$ at 
time $t$. In a system of $N$ individuals, 
opinions change out of binary interaction according to:
\begin{equation}
\label{eq:lccc}
\begin{aligned}
 o_i(t+1) &=& \lambda[o_i(t) + \epsilon o_j(t)] \\
 o_j(t+1) &=& \lambda[o_j(t) + \epsilon^\prime o_i(t)],
\end{aligned}
\end{equation}
where $\epsilon$, $\epsilon^\prime$ are randomly and independently drawn from an uniform 
distribution in $(0,1)$ and $\lambda$ is a parameter, interpreted as `conviction', the value of which lies between $0$ to $1$. 
In the above model~\cite{lallouache2010opinion} (LCCC model hereafter), everyone has the same 
value of conviction $\lambda$. In the above dynamics (Eq.~\ref{eq:lccc}), agent $i$ meets $j$, 
each keeping a fraction $\lambda$ of their own opinion and are also influenced by a random 
fraction of the other agent's opinion.
There are no conservation laws here for the opinions, but the opinions are bounded, i.e., $-1 
\le o_i(t) \le 1$. This bound is ensured by keeping the magnitude of the opinion values to $1$ even if Eq. (\ref{eq:lccc}) gives a higher magnitude. 
 The ordering in the system is measured by 
a quantity (order parameter) $O= |\sum_i o_i |/N$. 
Numerical simulations show that the multi-agent system (dynamics defined by 
Eqn.~(\ref{eq:lccc})) goes into either of the two possible phases: for any $\lambda \le 
\lambda_c$, $o_i=0$ $\forall i$, while for $\lambda > \lambda_c$,
$O>0$ and $O \to 1$ as $\lambda \to 1$, with $\lambda_c \approx 2/3$. 
Here $\lambda_c$ is the critical point of the phase transition.
The relaxation time, defined as the time to reach a stationary value of $O$ in 
time, diverges as $\tau \sim |\lambda-\lambda_c|^{-z}$ ($z \approx1.0\pm0.1$) when 
$\lambda \to \lambda_c$ on either side~\cite{lallouache2010opinion}. 
The order parameter near the critical point behaves 
as: $O \sim (\lambda - \lambda_c)^{\beta}$ with the order parameter exponent $\beta = 0.10 
\pm 0.01$~\cite{biswas2011phase}. 
A mean field calculation can be proposed for the fixed point $o^*$:
$o^*[1 - \lambda (1 + \langle \epsilon \rangle)]=0$, 
from which it easily follows that the critical point is $\lambda_c = 1 /(1+ \langle \epsilon 
\rangle)$ (where $\langle \ldots \rangle$ refers to average).
For uniform random distribution of $\epsilon$, $\langle \epsilon \rangle = 1/2$ and hence,
$\lambda_c = 2/3$. 
It was also noted that the underlying topology ($1d$, $2d$ or infinite range) 
has little effect on the critical point.
A map version of the model~\cite{lallouache2010opinion} has also been proposed:
$o(t+1) = \lambda (1+\epsilon(t)) o(t)$,
with $\epsilon(t)$  drawn randomly from a uniform distribution in $[0,1]$,
and $o(t)$ is still bounded in $[-1,+1]$ as before. The critical value $\lambda_c$
can be analytically shown to be $\exp[-(2\ln 2 -1)] \approx 
0.6796$~\cite{lallouache2010opinion}.

We introduce a noise term in the original equations (Eq.~\ref{eq:lccc}), such that the opinions 
now change out of binary interaction according to:
\begin{equation}
\label{eq:nlccc}
\begin{aligned}
 o_i(t+1) &=& \lambda[o_i(t) + \epsilon o_j(t)] + \zeta_i  \\
 o_j(t+1) &=& \lambda[o_j(t) + \epsilon^\prime o_i(t)] + \zeta_j,
\end{aligned}
\end{equation}
where $\zeta_i, \zeta_j$ are drawn randomly and independently from $(-\zeta,\zeta)$ with $|\zeta|\le 1$. Of course 
$\zeta=0$ corresponds to the LCCC model.
We observe that introducing the noise $\zeta$ destroys the active-absorbing nature of the 
phase transition. Specifically, in the LCCC model, the entire phase $\lambda < \lambda_c$ was 
characterized by null value of opinion for all agents, $o_i=0$ $\forall i$ and thus $O=0$ 
trivially. 
Any finite $\zeta$ remarkably changes this character, and $O \ne 0$ close to the critical point
(see Fig.~\ref{fig:nlc3M}b).
Another interesting observation is the change in the nature of the order parameter curve as 
the noise $\zeta$ increases. For small values of noise, the rate of change in $O$ is much 
faster near the critical point i.e., the transition is sharp, while this rate seems to decrease as the noise level increases 
(see Fig.~\ref{fig:nlc3M}a). This is because, the critical region is smaller, the closer the model is to the absorbing transition point. With an increased
noise parameter, the model is further away from the absorbing transition point and therefore has a wider critical range and hence show a relatively less sharp transition.
But as we shall see in the following discussions, the universality class of the transition is independent of the noise level.

\begin{figure}[t]
\includegraphics[width=17cm]{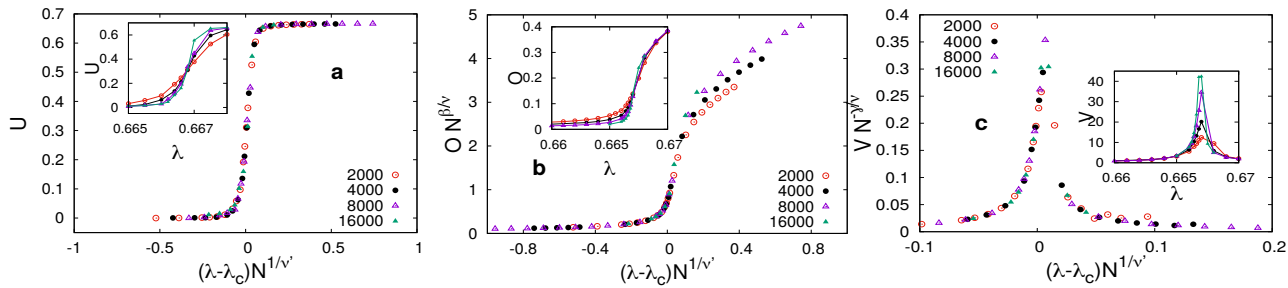}
\caption{Finite size scaling behavior for $\zeta=0.3$:
(a) Scaling collapse of Binder cumulant with $\lambda_c=0.6667 \pm 
0.0002$  estimated from the crossing for different sizes $N$ (inset). 
$\nu^\prime\equiv d\nu=2.00 \pm 0.01$ is estimated from the scaling collapse. 
Critical Binder cumulant $U^* = 0.30 \pm 0.01$.
(b) Scaling collapse of order parameter $O$ for $\beta = 0.50 \pm 0.01$.
Inset shows unscaled data for $O$ with $\lambda$.
(c) Scaling collapse of $V$ with $\gamma = 1.02 \pm 0.01$.
Inset shows unscaled data for $V$ with $\lambda$. 
 }
 \label{fig:0.3}
\end{figure}

Apart from the order parameter $O$, we also calculate the following quantities:
$V=N \left[ \langle O^2 \rangle - \langle O \rangle^2 \right]$ 
analogous to susceptibility per spin, and $U= 1- \frac{\langle O^4 
\rangle}{3\langle O^2 \rangle^2 }$, the fourth order Binder cumulant.
The order parameter $O$ behaves as $O \sim |\lambda-\lambda_c|^{\beta}$ and the susceptibility 
$V$ as $V \sim |\lambda-\lambda_c|^{-\gamma}$ near the critical point, i.e., for small values 
of $|\lambda-\lambda_c|$.
We apply finite-size scaling (FSS) theory to calculate the critical exponents. For large system 
sizes, we expect an asymptotic FSS behavior of the form
\begin{eqnarray}
 O= L^{-\beta/\nu} \mathcal{F}_O (x) \left[ 1+\ldots \right]\\
 V= L^{\gamma/\nu} \mathcal{F}_V (x) \left[ 1+\ldots \right],
\end{eqnarray}
where $\beta, \gamma$ are scaling exponents for the order parameter and the 
susceptibility respectively, $\cal{F}$ being the scaling functions with 
$x= (\lambda - \lambda_c)L^{1/\nu}$ as the scaling variable, $L$ being the linear dimension of 
the system. The dots in $[1 + \ldots]$ are the corrections to the scaling terms.
The crossing of the Binder cumulant is used to estimate $\lambda_c$ since Binder cumulant $U$ 
is independent of system size at the critical point $\lambda_c$, i.e., $U_L(\lambda_c)=U^*$. 
Scaling collapse of $U$ with $(\lambda - \lambda_c)L^{1/\nu}$ provides an estimate of critical 
exponent $\nu$. We study the infinite dimension (mean field) version  of the model, where any 
agent can interact with any other agent. In that case the linear dimension $L$ and system size 
$N$ are related as $L^d =N$, where $d$ will be the upper critical dimension. Hence,  
$L^{1/\nu}$  will be replaced by  $N^{1/\nu^\prime}$, where $\nu^\prime = \nu d$.

We estimate the critical point $\lambda_c$ for values of the noise parameter $\zeta$ in 
$(0:1)$, and thus get a phase diagram in the $\lambda-\zeta$ plane (Fig.~\ref{fig:nlc3M}c). 
The solid line is the critical line of $\lambda_c$ (or $\zeta_c$ if we had fixed the noise 
parameter and observed the transition by driving $\lambda$).
It seems that the critical point $\lambda_c$ increases very slowly by increasing $\zeta$ 
until $\zeta \simeq 0.4$, after which the rate of change is markedly faster. 

\begin{figure}[t]
\includegraphics[width=10cm]{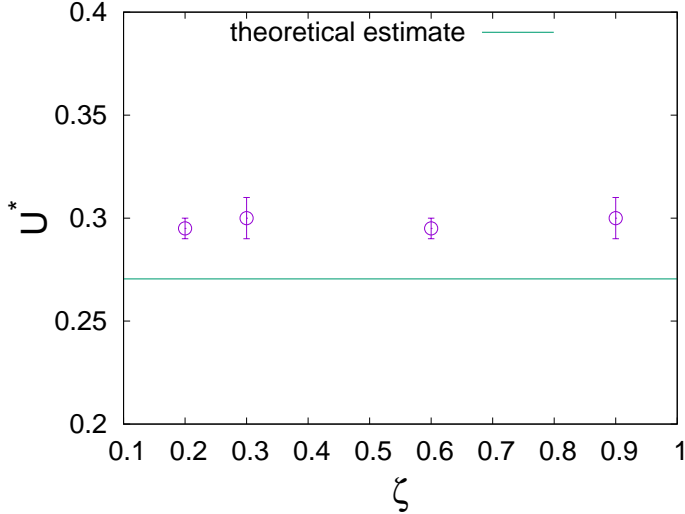}
\caption{ The variation of the critical Binder cumulant value ($U^*$) is shown for different noise parameters ($\zeta$). 
This value, known to indicate a universality class, is showing no systematic variation with $\zeta$. It is close to the 
theoretically estimated \cite{b_ising} value $\approx 0.27$ expected for the Ising model universality class in the mean field limit.
This also suggests that the kinetic exchange opinion model with finite noise amplitude belongs to the Ising universality class. 
 }
 \label{cr_bc}
\end{figure}

The value of the critical Binder cumulant ($U^*$) for all the noise amplitudes we studied ($\zeta=0.3,0.6,0.9$)
show a value close to $0.30$ (see Fig. \ref{cr_bc}), which is near the value ($\approx 0.27$) predicted  for the critical Binder cumulant of the Ising model (using field theoretic $\epsilon$ expansion \cite{b_ising}; see also the numerical estimate $\approx 0.30$ \cite{novotny}).
Numerical analysis here of the model gives $d\nu \simeq
2$, $\beta \simeq 1/2$ and $ \gamma \simeq 1$ practically for
all non-zero values of the annealed noise parameter
$\zeta$. It then implies from the hyper scaling relation
that the specific heat exponent $\alpha = 2$; $d\nu = 0$.
Then the Rushbrooke scaling relation $\alpha + 2\beta +
\gamma = 2$ gets satisfied. These  observations suggest
that the steady state statistics of the model belongs to
the Ising universality class for any amount of annealed
noise

\begin{figure}[t]
\includegraphics[width=17cm]{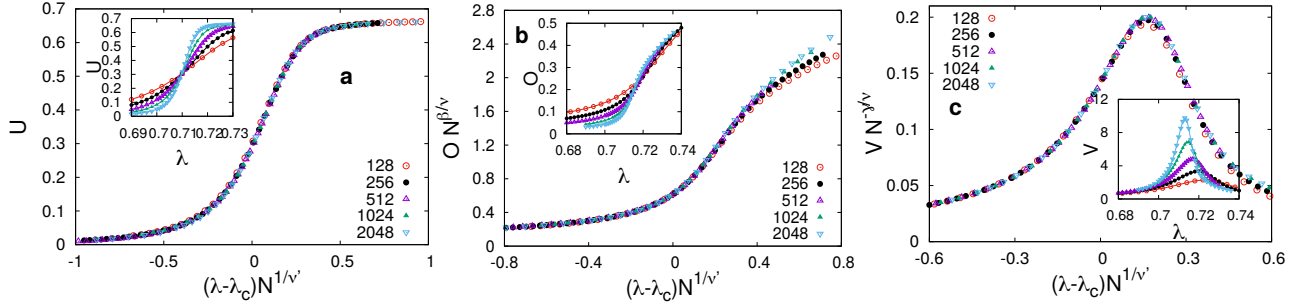}
\caption{Finite size scaling behavior for $\zeta=0.6$:
(a) Scaling collapse of Binder cumulant with $\lambda_c=0.7095 \pm 
0.0002$  estimated from the crossing for different sizes $N$ (inset). 
$\nu^\prime\equiv d\nu=2.00 \pm 0.01$ is estimated from the scaling collapse. 
Critical Binder cumulant $U^* = 0.295 \pm 0.005$.
(b) Scaling collapse of order parameter $O$ for $\beta = 0.43 \pm 0.01$.
Inset shows unscaled data for $O$ with $\lambda$.
(c) Scaling collapse of $V$ with $\gamma = 1.02 \pm 0.01$.
Inset shows unscaled data for $V$ with $\lambda$. 
 }
 \label{fig:0.6}
\end{figure}


\begin{figure}[t]
\includegraphics[width=17cm]{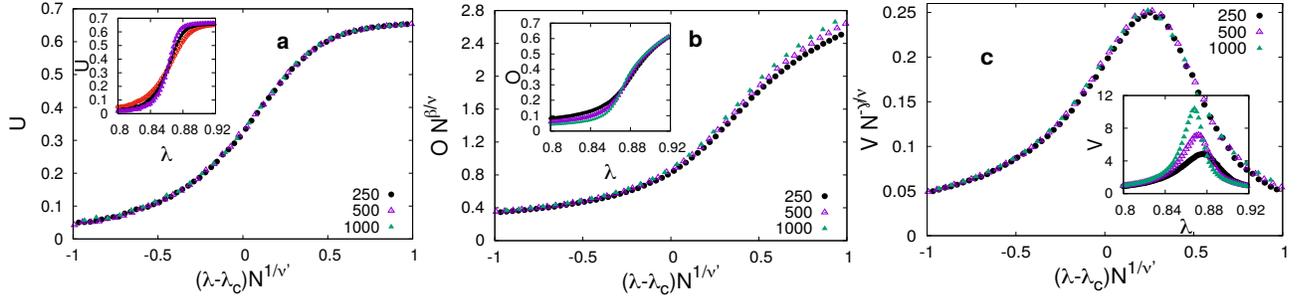}
\caption{Finite size scaling behavior for $\zeta=0.9$:
(a) Scaling collapse of Binder cumulant with $\lambda_c=0.8610 \pm 
0.0005$  estimated from the crossing for different sizes $N$ (inset). 
$\nu^\prime\equiv d\nu=1.96 \pm 0.01$ is estimated from the scaling collapse. 
Critical Binder cumulant $U^* = 0.30 \pm 0.01$.
(b) Scaling collapse of order parameter $O$ for $\beta = 0.50 \pm 0.01$.
Inset shows unscaled data for $O$ with $\lambda$.
(c) Scaling collapse of $V$ with $\gamma = 1.05 \pm 0.05$.
Inset shows unscaled data for $V$ with $\lambda$. 
 }
 \label{fig:0.9}
\end{figure}

\section{Kinetic exchange opinion model: Noise in the interaction}

In the infinite range interacting kinetic exchange opinion model introduced
by Biswas, Chatterjee and Sen \cite{biswas2012disorder}, the dynamical equation for the 
evolution of the individual opinion  $o_i(t)$  of the $i$-th agent at time
$t$, in presence of an external influence (denoted by a field term $h$),
can be extended (following \cite{biswas2012disorder}) to:
\begin{equation}
o_i(t+1)=o_i(t)+\mu_{ij}o_j(t)+h\cdot sgn(o_i(t)),
\end{equation}
where the agent-agent interaction term $\mu_{ij} = -1 $ with probability
$p$ and $1$ with probability $(1-p)$,  and the external influence term $|h|$, a stochastic variable,  is equal to $1$
with probability $q$ and equal to $0$ with probability $(1-q)$. It acts like a local reinforcement, triggered by external influences (e.g., consumption of news from favored but unreliable sources). The bound in the opinion values i.e., $|o_i|\le 1$ 
is enforced in the same way as before. However, the role of the conviction parameter is absent in this model. 
Here, the transition is driven by (even without external noise) negative interactions. This gave rise to 
an order-disorder transition with Ising critical exponents \cite{biswas2012disorder}, as opposed to ac active-absorbing transition \cite{lallouache2010opinion}.
Even without the last term, therefore, $o_i$ and $o_j$ could come closer together or move furher apart, depending on $\mu_{ij}$, which was not the case
for Eq. (\ref{eq:lccc}). The similarity, therefore, is in the binary interactive nature of the models and also the fact that the noisy variant shows Ising universality class, as we shall discuss.

If $f_1$, $f_{-1}$ and $f_0$ are the fraction of agents having opinion vales $1$, $-1$ and $0$ respectively, then in the steady state
\begin{equation}
f_0f_1(1-2p)+f_0f_{-1}(2p-1)+f_1f_{-1}(1-p)q+p(1-q)(f_{-1}^2-f_1^2)=0,
\label{probabilities}
\end{equation} 
where $f_1-f_{-1}=O$, the order parameter.
Further exact evaluation of the probabilities along this line turns out to be difficult. But 
one can write down a general polynomial expansion by collecting the $q$ dependent (linear) terms and
the $q$ independent terms, from an equation of the form Eq. (\ref{probabilities}). We can have
\begin{equation}
a+bO+cO^2+dO^3\dots=e\cdot q(1+\mathcal{O}(O)+\dots),
\label{expansion}
\end{equation}
where $a,b,c,d,e$ are constants.  

First, $a=0$, since we must have $O=0$ when $q=0$ and $p=p_c=1/4$. We must also have $c=0$, because 
$O(-h)=-O(h)$ (note that $e$ can depend on $h$, and thereby taking care of the sign on the right hand side).
Eq. (\ref{expansion}) could be compared with Eq. (3) of Ref. \cite{biswas2012disorder} in absence of $q$. The linear term
is present there, but not the cubic term. We argue that the cubic term is the only possibility beyond the
linear term and not the quadratic term, due to the reasons mentioned above. The cubic term, and possibly other higher order terms, will
appear in the equation when we take into account variations of the order parameter in two or more successive scatterings
(we only take one step scattering in Ref. \cite{biswas2012disorder}, but for the system to be in the steady state, we can take
the other steps in general). Stopping at the cubic term puts the requirement that $b/d \propto p-p_c$. 
Also, from Eq. (3) of Ref. \cite{biswas2012disorder}, it can be seen explicitly that the coefficient of the linear in $O$
term is proportional to $p-p_c$ (take $f_0=1/3$, then $p-f_0(1-p)$ becomes proportional to $p-p_c$). 
Combining these two observations suggest that $b \sim (p - p_c)$, and
in absence of the opinion influencing field ($e = 0$). Eq. (\ref{expansion})
then gives $O \sim b/d \sim (p-p_c)^{\beta}$, where $\beta$ = 1/2
(as already obtained in Ref. \cite{biswas2012disorder})
 
Finally, for $p\approx p_c$ keeping upto the linear terms in $O$ and $q$, which we take to be very small,
we have
\begin{equation}
bO\approx e\cdot q.
\end{equation} 

Then the susceptibility would be
\begin{equation}
\chi=\left.\frac{\partial O}{\partial q}\right|_{q\to 0}.
\end{equation}
Implying,
\begin{equation}
 \chi \sim \frac{1}{b} \sim (p-p_c)^{-\gamma},
\end{equation}
where $\gamma=1$, agreeing well with the numerical estimate in Ref. \cite{biswas2012disorder}.

Given that the order parameter exponent $\beta$, susceptibility
exponent $\gamma$ values obtained here and the numerically
estimated value of $d\nu = 2.0$ (in Ref. \cite{biswas2012disorder}), in this
long-range interacting version of the model  is confirmed to be
in the Ising universality class. However,  the dynamical
exponents (like persistence exponent), measured for lower dimensions, seem to differ from
the Ising class \cite{brexit}.

\section{Discussions}\label{sec:disc}

We have studied two variants of the infinite range interacting kinetic
exchange opinion model, namely that by Lallouache et al. \cite{lallouache2010opinion} (with an
added annealed noise to get continuous order-disorder transition, avoiding
the active-absorbing one)  and by Biswas, Chatterjee  and Sen \cite{biswas2012disorder},
using Monte Carlo simulations (analyzing finite size scaling, Binder cumulant
behavior, etc.) and scaling arguments (see Landau type expansion of the
steady state opinion probabilities in section 3). We show that at least in
the infinite range limit of interactions between the agents in these two
kinds of kinetic exchange models (of opinion formation dynamics), the
steady state statistics clearly belong  to the mean-field Ising universality class ($\beta = 1/2,
\gamma = 1$ and $d\nu = 2$, satisfying Rushbrooke scaling
$2\beta + \gamma = d\nu$, giving $\nu=1/2$ for upper critical
dimension $d =4$). Additionally, the value of the Binder cumulant at the critical point ($U^*$)
shows a value close to $0.30$ consistently for these models, which is close to the field theoretically estimated
value ($\approx 0.27$) for the Ising model in the mean field \cite{b_ising} and comparable with similar numerical
estimates \cite{novotny}. 
 The numerical results for lower dimensional
systems (see e.g., \cite{BCS2d3d}) also indicate the same. There could be interesting future directions in extending the
model for more realistic networks and/or models with diluted bonds (representing mutually non-interacting agents), to check if
the universality class changes.

It may be mentioned that this observation of Ising Universality class in
the  steady state statistics of some kinetic models help comprehending the
missing link connecting the observed Ising universality class of liquid
gas transition in the extended kinetic model of ideal gas, bypassing the
use of lattice-gas model.

\section*{Acknowledgments}
We are honored to have this opportunity to contribute in this special issue of Physica A
in memory of Prof Dietrich Stauffer, who pioneered in several interdisciplinary researches in statistical physics. 
The authors thank Parongama Sen for her valuable comments on the manuscript. 

\bibliographystyle{unsrt}
\bibliography{refmcc}

\end{document}